# Cloud Computing Concept and Roots

Bola Abimbola

*Abstract*—Cloud computing is a particular implementation of distributed computing. It inherited many properties of distributed computing such as scalability, reliability and distribution transparency. The transparency middle layer abstracts the underlying platform away from the end user. Virtualization technology is the foundation of Cloud computing. Virtual machine provides abstraction of the physical server resources and securely isolates different users in multi-tenant environment. To the Cloud services consumer, all the computing power and resources are accessed through high speed internet access by client platforms. This eliminates the cost to build and maintain local data center. Resource pooling and rapid elasticity are the main characters of Cloud computing. The scalability of Cloud computing comes from resources which can span multiple data centers and geographic regions. There is virtually no limitation on the amount of resources available from Cloud. New processing and storage resources can be added into the Cloud resource pool seamlessly.

## I. INTRODUCTION

ACCORDING to NIST, Cloud computing is a model where computing resources can be provisioned and release on-demand by consumer with self-service options and very little help from service providers [10]. The advancement of networking and other areas of computer hardware has made the Cloud computing model a reality. Amazon launched the first public accessible Cloud computing infrastructure with EC2 and S3 in 2006 [1]. This is a milestone in the Cloud computing era. Since then, the competition between major Cloud providers has resulted in major technology innovations and standardization initiatives. Today, the major public Cloud service providers are Amazon Web Services, Google Cloud Platform, Microsoft Azure, IBM Bluemix and Salesforce. This paper will discuss the Cloud computing roots then followed by its concept. The future of Cloud computing and recent development trend will also be discussed. Therefore, the paper will provide a comprehensive overview of what cloud computing is, how its services are provided and what are the different service models available for providing cloud computing services to customers according to their requirements.

## II. ROOTS OF CLOUD COMPUTING

While discussing about various cloud computing services and associate service delivery models, it is important to have a clear idea regarding roots of cloud computing to understand fundamental nature and characteristics of cloud computing that makes it different from traditional computing services [11]. It is important to identify that cloud computing integrates technologies from different areas of computing and networking and it is integration of these computing areas that makes it possible to deliver fundamental services offered by cloud computing technology. Analysis of these areas is essential to identify how cloud computing actually works and how it can be improved not only from a technological perspective but from a service perspective as well.

Cloud computing is a model developed from several other distributed computing research areas such as HPC, virtualization, utility computing and grid computing. These are some of the important technologies behind revolution of cloud computing and makes it possible to deliver high capacity processing power and computing storage over internet. Cloud computing has fundamentally changed the way computing services are delivered to the customers, but without advancement in these technologies, it was not possible. The cloud computing in its initial phase was not as diverse and as efficient as it is today, but over the years as new technologies have been introduced and existing technologies have been improved, it has revolutionized how computing services fundamentally work and delivered to customers as new service delivery models have been introduced to ensure access to cloud computing services is seamless and easy while ensuring that it is consistent with client requirements as well [12].

Cloud computing emerged because of the recent technology advancements in networking, storage, processor and software. The popularity of cloud computing has increased significantly since its initial market release and many vendors saw opportunity of marketing this technology by offering various cloud computing related services. Major IT companies quickly jumped into the Cloud computing market to develop and promote their new technologies. As the technology continued to become more advanced and popular, many companies has entered this huge market of cloud computing to offer their services with varying price options to choose from. Although there are many vendors in the market providing their services, only few large companies has been successful in establishing their dominance in the market.

One of the major reasons why cloud computing has become so popular among business and organizations is because it provides opportunity to access high quality computing without investing in complex infrastructure. The whole the idea of cloud computing service is to simply operations and management of computing services that organizations need to enhance their business processes.

Cloud computing also promises to simplify the resource management, fault tolerant, security and quality of services associated with large scale operations. Compare to grid computing, the dominant distributed computing model before Cloud model emerged, Cloud computing has clear advantages. There are no hardware differences in Cloud. Cloud provides a homogeneous computing environment where user has full control to OS with root access. There is no need to compile special version of source code in Cloud, because there are no more architecture differences in the Cloud computing resources [5].

Cloud computing is progressing towards the realization of utility computing [4]. Utility computing is a business model to meter the computing resource consumption. Computing power, storage spaces are packaged as services to be consumed like public utility [6]. All of this is made possible with the high speed Internet.

## III. Cloud Computing Concept

### A. Service Models

In order to ensure effective service delivery, it needs to consider proper service delivery model as for each service delivery model, type of services that is offered and how it is offered is different. However, choice for cloud service model depends on requirements and pricing preferences of different clients and there is no such best service delivery model. Therefore, before, choosing the service delivery model organizations need to analyze their requirements and select a particular service delivery models according to their requirements. Each service delivery models has its own benefits and limitations and therefore choice of service delivery model is essential while considering to invest in cloud computing for accessing computing resources and benefits offered by cloud computing technology as well.

There are three major service models of Cloud computing infrastructure as a service IaaS, platform as a service PaaS and software as a service SaaS [4].

IaaS gives consumer administrative privilege access to virtual machines along certain networking features to interconnect the virtual machines to internal and external services. Cloud service provider will manage the physical server hardware and networking. Consumers will manage everything else at the virtual machine, operating system and applications. Consumers do not need to worry about hardware failure because the high availability feature provided by virtualization [4]. Therefore, in this service delivery model, organizations are offered required infrastructure support that they will require to design and implement their applications and everything related to deployment and maintenance of applications are responsibility of the client. This type of service delivery model are helpful for organizations who have required expertise in design, deployment and maintenance of applications but do not want to invest in complex infrastructure design including maintenance of on-premise servers, storage and network equipment.

PaaS provides consumer with a development platform for their needs. Consumer will no longer have access to the operating system, but a set of libraries to use specific to the platform. The operating system is hidden from consumers. Consumers have to develop application to run on PaaS by following service provider guidelines. PaaS will provide scaling and high availability without consumer intervention. PaaS is less flexible than IaaS, and build on top of IaaS [4].

SaaS is built on top of PaaS and IaaS. The software service provider is responsible for everything except the client program. Consumers can access the software via Internet with a subscription model. SaaS is normally offsite and has multiple tenants. Multiple consumers' data is processed, stored and managed by SaaS. As a result, SaaS is might not suitable for certain industries like government and financial services [4].

As already mentioned that each service delivery model has its own benefits and limitations, it requires organization to decide which service delivery model address their need better and suits their business requirements. An extensive market research is required before purchasing cloud computing services and sometime vendors can also offer insight in this regard. However, it is responsibility of organizations that have complete understanding of scalability and flexibility offered by each service delivery model so that investment in cloud computing is effective and efficient as well.

### B. Virtualization and Fault Tolerant

As in cloud computing model depending on whether it is private cloud or private cloud model, many organizations are accessing computing resources. If computing resources like servers are not properly managed, then it might affect availability of resources. It is extremely important to ensure that servers are available whenever required by users so that it provides a seamless computing experience.

Virtualization is the enabling technology for most Cloud platforms. It is used in data centers to consolidate underutilized servers. Through consolidation, virtualization can create the resource pool for the Cloud environment. Virtualization of storage and networking are also important technologies for Cloud computing [9]. Virtual machines can be created instantly from templates or image files. By using shared storage, live migration and fault tolerant features of virtualization hypervisor, virtual machine can be configured with high availability to ensure SLA [2].

### C. Elasticity Through Resource Pooling

Computing resources in the Cloud is presented to the consumer as a giant pool. The resource pool contains computing resources from multiple data centers. There is virtually no limit on how big this pool can be. Cloud service providers usually describe each pool as regions and zones. The datacenter layout or hardware server configurations are completely hidden for the consumer. Consumer has no knowledge where exactly the resources come from [7]. Elasticity means the consumer can provision and release resource as needed. Consumer does not have to have any up front commitment on any computing resources. When there is a need for resource, it can be provisioned immediately.

One of the main reasons for organizations to adopt cloud computing services is that they do not have to worry about resource availability. However, demand for resources depends on workload and cannot be decided accurately upfront. Cloud computing offers this facility of offering resource on demand as resources are delivered only when required. Delivering resources according to demand ensures that resources are wasted and can be utilized elsewhere when not required. It ensures resource availability to every user. Therefore, elasticity is an important concept in context of cloud computing for optimization of resources and enhancing quality of cloud computing services.

### D. Self-Serve and On-Demand

Cloud computing providers offer all types of services to consumers on-demand with pay as you go business model. Cloud consumers save cost since they only pay for what they use. There is no need to purchase hardware upfront along with any human resource required to setup and maintain the onsite facility. This is especially valuable to certain situation where demand can peek and fall from time to time. Combined with automated scaling, on-demand elasticity not only saves cost, but also increase the client system robustness and responsiveness [9]. Therefore, cloud computing provides cost efficient and powerful computing solutions that helps organizations in managing their business requirements as well.

## IV. CONCLUSION

Cloud computing market will continue to grow driven by the migration from traditional datacenter to the Cloud. Many business data processing such as big data is leveraging the power of Cloud. Cloud computing has made it easier for organizations to collaborate on global scale as data can be shared and accessed from anywhere around the world and therefore, increasing productivity of organizations. However, one of the main concerns regarding cloud computing is data security specially when cyber security issues are increasing which require strong measures against it for restricting unauthorized access to important business and organizational data as well. Therefore, it is challenging for cloud service vendors to ensure data security while offering benefits of cloud computing.

Although Cloud security is one of the biggest issues for many organizations to adopt Cloud computing, it is clear that major Cloud platforms are taking security seriously and working hard to comply with international security compliance policies. Various advance technologies like artificial intelligence, machine learning are being implemented for optimizing data security management. Cloud computing is still in its early life, there are active research and development around it. It will not only improve cloud services but it will also enhance data security as well.